\documentclass{ptephy_v1}

\preprintnumber{XXXX-XXXX} 
\usepackage{amsthm}
\usepackage{bm,latexsym,amsmath,amssymb,amsfonts}
\usepackage{graphics}
\usepackage{ulem}

\begin{document}

\title{Massive fermion between two parallel chiral plates}

\author{Ar Rohim$^{1,2}$, Apriadi Salim Adam$^{2}$,  and Kazuhiro Yamamoto$^{1,3}$}
\affil{$^1$Department of Physics, Kyushu University, 
  Fukuoka,  819-0395, Japan
 \\
 $^2$Research Center for Quantum Physics, National Research and Innovation Agency (BRIN), South Tangerang 15314, Indonesia
  \\
 $^3$Research Center for Advanced Particle Physics, Kyushu University, 
 Fukuoka 819-0395, Japan}

\begin{abstract}
We study the system of a massive fermion  field confined between two parallel plates, where the properties of both plates are discussed under chiral MIT boundary conditions. 
We investigate the effects of the chiral angle on the Casimir energy for a massive fermion field with the general momentum. 
We find that the Casimir energy as a function of the chiral angle is generally symmetric, and the attractive Casimir force in the chiral case is stronger than that in the nonchiral case. In addition, we investigate the approximate Casimir energy for light and heavy mass cases. The behavior of the discrete momentum and changes of spin orientation are also discussed.
\end{abstract}

\subjectindex{A64, B30, B69}

\maketitle

\section{Introduction}
The Casimir effect for two parallel conducting plates in a vacuum state that generates an attractive force was first discussed in Ref.~\cite{Casimir1948}. 
 In 1958, an experimental measurement on such a force was performed with rough precision \cite{Sparnaay:1958wg}, and the precision of the measurement has since been increased \cite{Lamoreaux:1996wh, Mohideen:1998iz, Roy:1999dx, Bressi:2002fr} (see also Refs.~\cite{Bordag:2001qi, Onofrio:2006mq} for review).
The theoretical discussion was extended for various models (see, e.g., Refs.~\cite{Lutken1984, Zahed1984, Valuyan2009, Elizalde:2011cy, Fosco:2008vn, Bellucci:2009jr, Seyedzahedi2010, Oikonomou:2009zr, Fosco2022, Shahkarami2011, Moazzemi2007, DePaola1999, Boyer1968, Mobassem:2014jma, Edery:2006td, Cruz:2017kfo, Erdas:2021xvv, Erdas:2010mz, Ambjorn:1981xw}). The Casimir effect is a manifestation of the quantum field theory with appropriate boundary conditions. 
Thus, the choice of the boundary condition plays an important role in the Casimir effect.   The essential aspect when discussing the Casimir effect is not only placed on the type of boundary condition involved but also the type of quantum field.  For the case of the scalar field, the variant of the Dirichlet and Neumann boundary conditions are frequently used in the literature \cite{Valuyan2009, Mobassem:2014jma,Moazzemi2007, Edery:2006td, Cruz:2017kfo, Erdas:2021xvv}. For the fermion field, however, these two boundary conditions cannot be applied \cite{Ambjorn:1981xw}. Instead, one may use alternative boundary conditions, e.g. a bag boundary \cite{Lutken1984, Saghian2012, Oikonomou:2009zr, Elizalde:2011cy, Fosco2022, Cruz2018, Elizalde2002, Bellucci2009, Zahed1984, DePaola1999, Erdas:2010mz, Ambjorn:1981xw}.

Discussing the boundary condition for a Dirac field is nontrivial because the Dirac equation is a first-order differential equation. To discuss the Casimir effect for the Dirac field, several authors \cite{Saghian2012, Cruz2018, Elizalde2002, Bellucci2009} have used the boundary condition in the MIT bag model \cite{Chodos11974, Chodos21974, KJohnson1975} (see also Refs.~\cite{Alberto1996, Alberto2011} for confinement system); this guarantees the vanishing flux or the normal probability density at the boundary surface. However, this boundary condition leads to a discontinuity of the axial-vector current at the boundary surface that breaks its chiral symmetry. An alternative way to address this issue is to introduce the chiral bag model in the presence of the pion field \cite{Chodos1975}.

A more general form of the boundary condition in the MIT bag model that includes the chiral angle is the so-called chiral MIT boundary conditions \cite{Theberge1980, Lutken1984, Jaffe1989}. Using this boundary, one can investigate the interaction between the particle and boundary surface, which  may change the spin orientation depending on the chiral angle \cite{Nicolaevici2017}. Thus, the roles of the chiral angle in boundary conditions for a Dirac field may give essential features (e.g. Refs.~\cite{Ambrus:2015lfr, Chernodub:2016kxh, Chernodub:2017ref, Chernodub:2017mvp, Rohim2021}).  Ref.~\cite{Rohim2021} showed that the particle's energy in the confinement system also depends on the chiral angle. There is another general form of the boundary condition in the MIT bag model, the self-adjoint boundary condition, which was used in Ref.~\cite{Sitenko:2014kza} to discuss the Casimir effect by including the background magnetic field (see also Ref.~\cite{Donaire}).

This study investigates the Casimir effect of a Dirac field confined between two parallel plates. The properties of both plates are described by chiral MIT boundary conditions \cite{Lutken1984}.  
 We propose the general solution for the Dirac equation in such a system following the arguments in Refs.~\cite{Alberto1996, Alberto2011, Nicolaevici2017}. We write the mass of a particle as a function of position and include an analysis of the spin orientation by distinguishing its field components. Along with the mentioned procedure, we discuss not only the Casimir energy but also the Casimir pressure. Compared with Ref.~\cite{Lutken1984}, where the authors applied the chiral MIT boundary conditions for the massless case (see also Refs.~\cite{Oikonomou:2009zr}), in this paper we apply the boundary conditions for the case of a massive fermion field.  We also investigate how the spin orientation changes under the interaction between the field and the boundary surface of the plates. Our detailed setup of the boundary condition in the first plate differs from that of Ref.~\cite{Lutken1984}, where the author set the specific value of the chiral angle for the first plate and took a general chiral angle for the second plate. In our setup, we set the chiral angle at both plates with the same general value.  From the viewpoint of the boundary conditions type, the present paper is an extension of the earlier study on the Casimir energy by the authors in Ref.~\cite{Saghian2012}, in which they discussed the nonchiral case.

 In this paper,  we also investigate the energy gap between two states under the effect of the chiral angle. Compared to the work in Ref.~\cite{Beneventano:2010ky} where the authors used local boundary conditions, in this study we compute such an energy gap derived using chiral MIT boundary conditions. Thus, we can address the effect of the chiral angle on the electron transport in material such as graphene nanoribbons \cite{MYHan2007, YMLin2008, Han2010}.  We also expect that the present study could be applied to nanotube under the chiral MIT boundary conditions. The application to such a system under the nonchiral boundary has been done previously by Ref.~\cite{Bellucci2009} (c.f., Ref.~\cite{Bellucci:2009jr}).

 The rest of this paper is organized as follows. In Sec.~\ref{PhysicalSetup}, we introduce the general setup for our model. In Sec.~\ref{GeneralFeatures}, we discuss the features under the chiral MIT boundary conditions focusing on the discrete momenta and change of spin orientations. In Sec.~\ref{CasimirEnergy}, we investigate the Casimir effect of a massive fermion under this boundary condition.  Section \ref{Summary} is devoted to our summary. In Appendix~\ref{detaildiscretemomenta}, we provide the complementary derivations for discrete momenta. Throughout this paper, we will use natural units $c=\hbar=1$.

\section{Physical setup}
\label{PhysicalSetup}

\begin{figure}[tbp]
\centering 
\includegraphics[width=.6\textwidth]{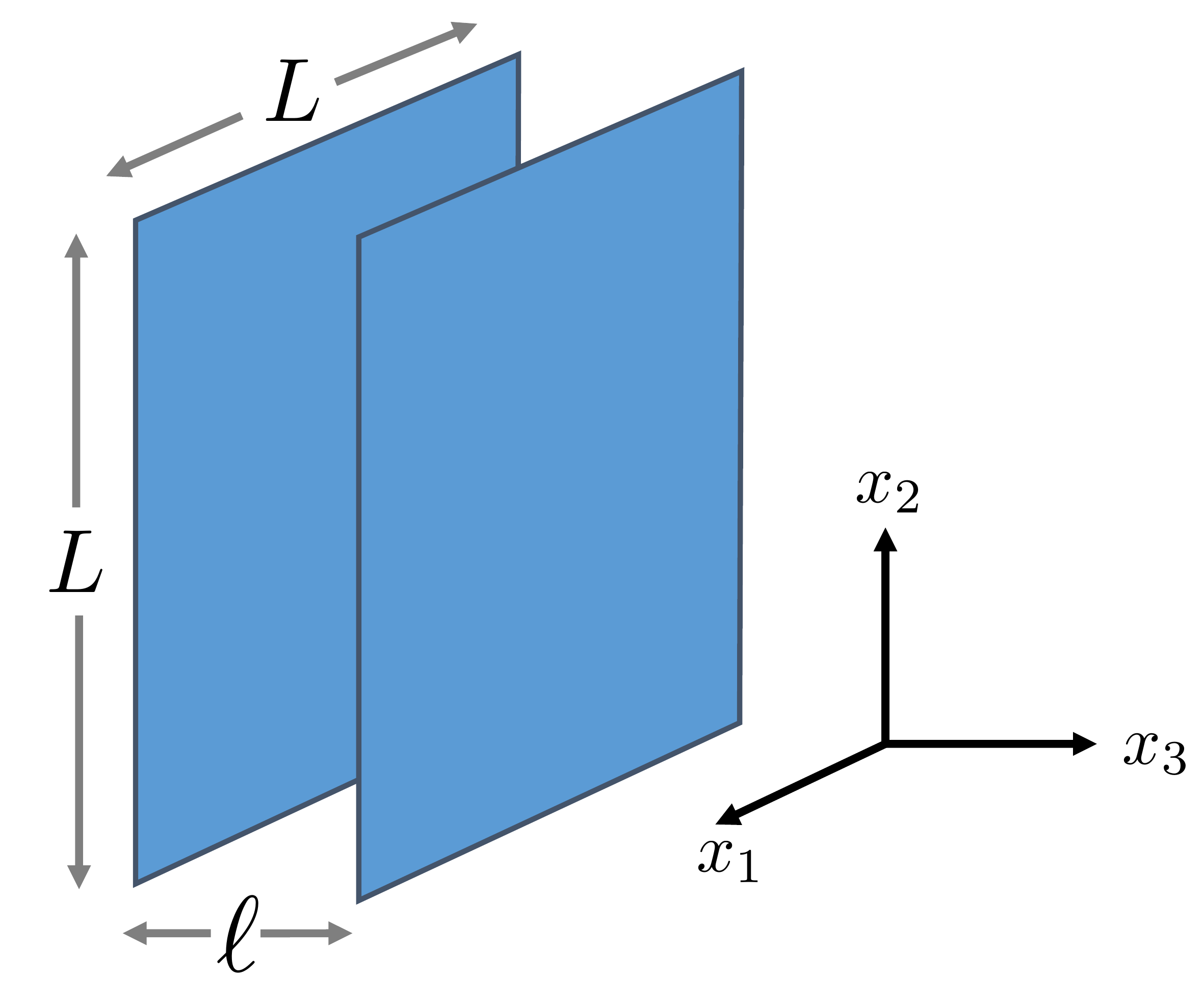}
\caption{\label{setup} Physical setup of a Dirac fermion confined between two parallel plates. We use $L$ and $\ell$ to represent the plate's size and distance, respectively. In the present paper, 
we assume the limit $L\rightarrow \infty$ approximately.}
\end{figure}

We consider a free massive Dirac fermion confined between two parallel plates in $3+1$-dimensional Minkowski spacetime background. The first plate is placed at $x_3=0$ while the second one is placed at $x_3=\ell$ (see Fig.~\ref{setup}).
Both plates are in parallel with $(x_1,x_2)$-plane. 
In such a system, the action of a Dirac field with mass $m$ is given by
\begin{eqnarray}
S=\int d^4x \bar \Psi(i\gamma^\mu \partial_\mu-m) \Psi,
\label{DiracAction}
\end{eqnarray}
where $\bar \Psi\equiv\Psi^\dagger \gamma^0$ is the Dirac adjoint and $\gamma^\mu$ are the gamma matrices 
in the Dirac representation given by 
\begin{eqnarray}
 \gamma^0=  \begin{pmatrix}
I & 0\\
0 &-I
\end{pmatrix}
~~{\rm and}~~
 \gamma^i=  \begin{pmatrix}
o & \sigma_i\\
-\sigma_i &0
\end{pmatrix},
~~ i=1,2,3~, 
\end{eqnarray}
 with $I$ is $2\times 2$ identity matrix and $\sigma_i$ are Pauli matrices. The above gamma matrices satisfy anticommutation relation $\lbrace \gamma^\mu,\gamma^\nu\rbrace=2\eta^{\mu\nu}$ with $\eta^{\mu\nu}={\rm diag.}(1,-1,-1,-1)$. 
Taking the variation of the action in  Eq.~(\ref{DiracAction}) leads to the following Dirac equation, 
\begin{eqnarray}
(i\gamma^\mu \partial_\mu-m) \Psi=0. 
\label{DiracEq}
\end{eqnarray}

To propose the specific form of the Dirac field in the region between two parallel plates that satisfies the Dirac equation (\ref{DiracEq}), we follow the arguments used in Refs.~\cite{Alberto1996, Alberto2011, Nicolaevici2017} as follows. (i) The particle mass depends on its position, which is originally described by the MIT bag model for hadron \cite{Theberge1980, Lutken1984, Jaffe1989}. Namely, inside the region between two parallel plates, the mass of a Dirac field is finite and becomes infinite at both plates \cite{Alberto1996, Alberto2011}. Under this condition, the Dirac field outside the confinement area vanishes. (ii) The form of a Dirac field consists of two-component fields associated with their spin orientations \cite{Nicolaevici2017}. Based on the  arguments above, the proposal for a positive frequency of the massive Dirac field in the region between two parallel plates is given by 
\begin{eqnarray}
&&\psi^{(+)}_{{\bm k}s}(t,{\bm x})=e^{-iE t} \psi^{(+)}_{{\bm k} s}({\bm x}),
\label{TotalDiracwavetime}\\
&&\psi^{(+)}_{{\bm k} s}({\bm x})= B
  \begin{pmatrix}\chi_R\\
r\hat {\bm k}_R\cdot{\bm \sigma} \chi_R
\end{pmatrix} e^{i{\bm k}_R \cdot {\bm x}}+C
  \begin{pmatrix}\chi_L\\
r\hat {\bm k}_L\cdot{\bm \sigma} \chi_L
\end{pmatrix} e^{i{\bm k}_L \cdot {\bm x}},
\label{TotalDiracwave}
 \end{eqnarray}
 where $B$ and $C$ are complex coefficients,   $s$ represents the spin orientation, $r=|{\bm k_{R(L)}}|/(m+E)$, and $E = \sqrt{k^2_1+k^2_2+k^2_3+m^2}$ is the energy  of a Dirac field.  The two-component spinors  $\chi_{R(L)}$  satisfy the normalized condition of  $\chi^\dagger_{R(L)}\chi_{R(L)}=1$ with subscripts $R$ and $L$ corresponding to the right and left of the Dirac field components, respectively. We also use the notations ${\bm k}_R=(k_1,k_2,k_3)$ and ${\bm k}_L=(k_1,k_2,-k_3)$  to represent the spatial momentum of the right- and left-moving field components.  Note that we have distinguished the two-component spinor for each component of the Dirac field in Eq.~(\ref{TotalDiracwave}) because their spin orientations may depend on the boundary condition 
\cite{Nicolaevici2017}. The corresponding form of the negative frequency for the massive Dirac field can be obtained by taking the charge conjugation of the  above positive-frequency Dirac field as
\begin{eqnarray}
\psi^{(-)}_{{\bm k}s}(t,{\bm x})=\psi^{(+)\rm C}_{{\bm k}s}(t,{\bm x})=i\gamma^2\psi^{(+)*}_{{\bm k}s}(t,{\bm x}).
\end{eqnarray}

In our model, the properties of both plates are described by chiral MIT boundary conditions given as  \cite{Lutken1984}
  \begin{eqnarray}
	iN_\mu\gamma^\mu\psi
	=e^{-i\Theta\gamma^5}\psi,
	\label{chiralMITboundaryconditions}
\end{eqnarray}
where $N_\mu$ is an inward normal unit four-vector perpendicular to the boundary surface, $\Theta\in [0,2\pi)$ denotes the chiral angle, and $\gamma^5\equiv i\gamma^0\gamma^1\gamma^2\gamma^3$. The above boundary condition guarantees the vanishing of the normal probability current density at the boundary surface  for any chiral angles \cite{Jaffe1989} 
\begin{eqnarray}
N_\mu J^\mu(\equiv N_\mu \bar \psi\gamma^\mu \psi)=0. 
\end{eqnarray}
In the nonchiral case $\Theta=0$, the boundary condition (\ref{chiralMITboundaryconditions}) reduces to that given in the MIT bag model.

\section{Features under boundary conditions}
\label{GeneralFeatures}

In this section, we investigate two features of a massive fermion field confined between two parallel plates under chiral MIT boundary conditions \cite{Lutken1984}. Namely, we discuss how the boundary condition affects the structure of the discrete momenta and the changes of the spin orientation following the procedure in Refs.~\cite{Nicolaevici2017, Rohim2021}. However, our system proceeds with the general momentum (see also Refs.~\cite{Saghian2012, Cruz2018, Bellucci2009}). 

\subsection{Discrete momenta} 
The boundary condition at the first plate $x_3=0$ is given by 
\begin{eqnarray}
i\gamma^3\psi|_{x_3=0} =e^{-i\Theta\gamma^5}\psi |_{x_3=0}, 
\end{eqnarray}
which is obtained from Eq.~(\ref{chiralMITboundaryconditions}) with the inward normal unit four-vector given as $N_\mu=(0,0,0,1)$. 
It can be rewritten into two equivalent equations as follows
\begin{eqnarray}
&&i(\sigma_3+\sin\Theta I)\chi_2|_{ x_3=0}-\cos\Theta\chi_1|_{ x_3=0}=0,
\label{chiral10}\\
&&i(\sigma_3-\sin\Theta I)\chi_1|_{ x_3=0}+\cos\Theta\chi_2|_{ x_3=0}=0,
\label{chiral20}
\end{eqnarray}
where $\chi_1$ and $\chi_2$ are the upper and lower two-components Dirac field, 
respectively. Applying the boundary conditions (\ref{chiral10}) or (\ref{chiral20}) to the positive frequency Dirac field (\ref{TotalDiracwave}), we obtain the relation of coefficients $B$ and $C$ at the first plate as 
\begin{eqnarray}
\big [i(\sigma_3+\sin\Theta I) r\hat {\bm k}_R\cdot {\bm \sigma}-\cos\Theta I\big] B \chi_R =- \big [i(\sigma_3+\sin\Theta I) r\hat {\bm k}_L\cdot {\bm \sigma}-\cos\Theta I\big] C \chi_L.
\label{relation01}
\end{eqnarray}
Below, we show that this relation is useful for investigating the discrete momenta and the changes of the spin orientation at the first plate.

We next employ the boundary condition at the second plate to know the behavior of discrete momenta. At the surface of the second plate $ x_3=\ell$, the inward normal unit four-vector reads  $N_\mu=(0,0,0,-1)$. Then, the corresponding boundary condition is given by
\begin{eqnarray}
-i\gamma^3\psi|_{x_3=\ell} =e^{-i\Theta\gamma^5}\psi |_{x_3=\ell}, 
\end{eqnarray}
which leads to the following two equivalent  equations
\begin{eqnarray}
&&i(-\sigma_3+\sin\Theta I)\chi_2|_{ x_3=\ell}-\cos\Theta\chi_1|_{ x_3=\ell}=0,
\label{chiral1L}\\
&&i(-\sigma_3-\sin\Theta I)\chi_1|_{ x_3=\ell}+\cos\Theta\chi_2|_{ x_3=\ell}=0.
\label{chiral2L}
\end{eqnarray}
Applying boundary conditions~(\ref{chiral1L}) or (\ref{chiral2L}) to the positive-frequency Dirac field (\ref{TotalDiracwave}), we have the relation as follows
\begin{align}
\big [i(-\sigma_3+\sin\Theta I) r\hat {\bm k}_R\cdot {\bm \sigma}-\cos\Theta I\big] e^{ik_3\ell} B \chi_R =- \big [i(-\sigma_3+\sin\Theta I) r\hat {\bm k}_L\cdot {\bm \sigma}-\cos\Theta I\big] e^{-ik_3\ell} C \chi_L. 
\label{relationL1}
\end{align}
Note that two-component Dirac spinors $\chi_R$ and $\chi_L$ in Eq.~(\ref{relationL1}) will be the same as those in Eq.~(\ref{relation01}) when the spin orientations are consistently reflected; namely, the reflected spin orientation at the second plate is the same as the incident spin orientation at the first plate (see Ref.~\cite{Rohim2021} for one-dimensional case).  

Utilizing relations given in Eqs.~(\ref{relation01}) and (\ref{relationL1}), we are able to derive the constraint for the momentum as follows  
\begin{eqnarray}
m\ell\cos\Theta\sin(k_3 \ell)+k_3\ell\cos(k_3\ell)= 0.
\label{DiscreteMomenta}
\end{eqnarray}
For the detailed derivation, see Appendix~\ref{detaildiscretemomenta}. The solution for Eq.~(\ref{DiscreteMomenta}) shows that the momentum $k_3$ must be discrete and
depends on the chiral angle.  It also shows that in our system, momenta $k_1$ and $k_2$ remain because the boundary conditions appear only at the $x_3$-axis in parallel with $(x,y)$-plane, as mentioned in the previous section.

Next, we define $k'_{3n}\equiv k_3\ell$ with $n=1,2,3,\cdots$ to denote the solution for Eq.~(\ref{DiscreteMomenta}). In comparison to the solution for the nonchiral case, we here have a factor of $\cos\Theta$ that contributes to determining the structure of the discrete momenta. However, 
when the chiral angle takes values as $\Theta=\pi/2, 3\pi/2$, the discrete momenta  have a nontrivial solution for all mass $m$ as follows \cite{Rohim2021},
\begin{eqnarray}
k'_{3n}={(2n-1)\pi\over 2},
\label{solnontrivialK3}
\end{eqnarray}
which is the same as the discrete momenta solution in the massless case $m=0$. 

As has been discussed in the previous work \cite{Rohim2021}, there are two  cases when analyzing the solution for  Eq.~(\ref{DiscreteMomenta}). 
For the case of light mass 
$m'(\equiv m\ell)\ll 1$, the solution of the discrete momenta is approximately given by
 \begin{eqnarray}
 k'_{3n, l}\simeq{(2n-1)\pi\over 2}+ {2 m'\cos\Theta\over (2n-1)\pi}.
 \end{eqnarray}
We note that the first term of the above solution covers the solution of Eq.~(\ref{solnontrivialK3}) while the second term gives the correction to the discrete momenta $k'_{3n}$ 
depending on the mass $m$ and the chiral angle $\Theta$. 
Whereas  in the case of heavy mass $m'\gg 1$, the solution for discrete momenta is approximately given by 
 \begin{eqnarray}
 k'_{3n, h}\simeq n\pi-{n\pi\over m'\cos\Theta}. 
 \label{k3nh}
 \end{eqnarray}
In Eq.~\eqref{k3nh}, the first term covers the discrete momenta for the  Schr\"{o}dinger equation in an infinite potential well under the Dirichlet boundary condition \cite{Rohim2021}. Since $m'$ depends on $\ell$, the case of $m'\ll 1$ and $m'\gg 1$  correspond to small and large distances between two plates, respectively. In other words, one can write the distance between two plates as a function of the Compton wavelength that determines whether the confinement system approaches ultra- or non-relativistic limits \cite{Alberto1996, Alberto2011}.

\subsection{Change of spin orientation  and energy gap}

The two-component spinor $\chi_{R}$ can be connected to  $\chi_{L}$ using a rotation operator in spin space, which is determined by the boundary condition. 
Once we know the structure of one of them, we will obtain complete information on the spin orientations for both Dirac field components (\ref{TotalDiracwavetime}) and  (\ref{TotalDiracwave}). 
The rotation operator in spin space is given by \cite{Nicolaevici2017}
 \begin{eqnarray}
 {\cal U} =e^{i\varsigma }\left[\cos({\varphi \over 2})I-i\sin({\varphi \over 2})
   \hat {\bm n} \cdot{\bm \sigma}\right],
   \label{rotoperatorspinspace}
  \end{eqnarray}
where $e^{i\varsigma }$ denotes a pure phase, $\varphi $ is the rotation angle, and $
 \hat {\bm n} $ represents the unit rotation axis generated by the reflection with the plate.

At the first plate, from the relation given in Eq.~(\ref{relation01}),  we have $\chi_R={\cal U}^{(1)}_R \chi_L$
with ${\cal U}^{(1)}_R$ is the rotation operator in spin space generated by the reflection with the first plate given as 
\begin{align}
{\cal U}^{(1)}_{R}={C\over B}{\cos\Theta \big[(k'^2_{3n}/\ell^2+m E_n+m^2)I+i(k'_{3n}/\ell)(k_2\sigma_1-k_1 \sigma_2+(m+E_n)\tan\Theta\sigma_3)\big]\over  (m+E_n)(ik'_{3n}/\ell-m\cos\Theta)},
\label{rotU1}
\end{align}
 where we have used the discrete momenta $k'_{3n}$ and the eigen energies
 \begin{eqnarray}
E_n=\sqrt{k^2_1+k^2_2+\bigg({k'_{3n}\over \ell}\bigg)^2+m^2}.
\label{Discreteenergy}
\end{eqnarray}
Taking the correspondence between the obtained rotation operator (\ref{rotU1}) and the general formulation in Eq.~(\ref{rotoperatorspinspace}), we have
\begin{eqnarray}
&&e^{i\varsigma^{(1)} }\cos({\varphi^{(1)} \over 2})I
={C\over B}{\cos\Theta (k'^2_{3n}/\ell^2+m E_n+m^2)I\over  (m+E_n)(ik'_{3n}/\ell-m\cos\Theta)}
,\\
&&e^{i\varsigma ^{(1)}}\sin({\varphi^{(1)} \over 2})\hat {\bm n}^{(1)} \cdot{\bm \sigma}
=-{C\over B}{\cos\Theta (k'_{3n}/\ell) \big[ k_2\sigma_1-k_1 \sigma_2+(m+E_n)\tan\Theta\sigma_3\big]\over  (m+E_n)(ik'_{3n}/\ell-m\cos\Theta)}
,
\end{eqnarray}
which lead to the expression of the  
rotation angle and its rotation axis at the first plate as follows\footnote{The rotation angle and rotation axis can also be written as $\tan({\varphi^{(1)}/2})=-\tan \Theta (k'_{3n}/\ell)\sqrt{2m^2+2mE_n+k'^2_{3n}/\ell^2
+{\bm k}^2_\perp \cot^2\Theta}/ (k'^2_{3n}/\ell^2+m E_n+m^2)$ and $\hat {\bm n}^{(1)}={k_2\hat x_1-k_1 \hat x_2+(m+E_n)\tan\Theta\hat x_3 /(\tan\Theta \sqrt{2m^2+2mE_n+k'^2_{3n}/\ell^2
+{\bm k}^2_\perp\cot^2\Theta}})$, respectively. These expressions reduce to those in  Ref.~\cite{Rohim2021} in the case of suppressed perpendicular momentum ${\bm k}_\perp=0$. As has been mentioned in Ref.~\cite{Rohim2021}, one may choose the opposite sign of the above rotation axis. As the consequence, the rotation angle has an opposite sign to the above one. This condition holds not only  for the first plate but also the second one.}
\begin{eqnarray}
&&\tan({\varphi^{(1)}\over 2})=-{(k'_{3n}/\ell)\sqrt{(2m^2+2mE_n+k'^2_{3n}/\ell^2
)\tan^2\Theta+{\bm k}^2_\perp\sec^2\Theta}\over  (k'^2_{3n}/\ell^2+m E_n+m^2)},
\label{rotangle1}\\
&&\hat {\bm n}^{(1)}={k_2\hat x_1-k_1 \hat x_2+(m+E_n)\tan\Theta\hat x_3 \over \sqrt{(2m^2+2mE_n+k'^2_{3n}/\ell^2
)\tan^2\Theta+{\bm k}^2_\perp\sec^2\Theta}},
\label{rotaxis1}
\end{eqnarray}
respectively  and ${\bm k}_\perp(\equiv \sqrt{k^2_1+k^2_2})$ is the perpendicular momentum to the normal surface of the plates. It can be seen that the roles of the pure phase $e^{i\varsigma ^{(1)}}$ can be cancelled out from the rotation angle and its axis. 
In the case of ${\bm k}_\perp\neq 0$, the reflected spin orientation changes for arbitrary chiral angles (both chiral and nonchiral cases). However, for the case of ${\bm k}_\perp=0$, it does not change for $\Theta=0, \pi$ (see, for example, Refs.~\cite{Cruz2018, Bellucci2009} for the discussion on the relation of $\chi_R$ and $\chi_L$ under nonchiral boundary condition, where  $B=C$ was used).

At the second plate, from the relation provided in Eq.~(\ref{relationL1}), we have $\chi_L={\cal U}^{(2)}_L \chi_R,$ where ${\cal U}^{(2)}_L$ is the rotation operator generated by the reflection with the second plate given by
\begin{align}
{\cal U}^{(2)}_{L}={B e^{2ik'_{3n}}\over C}
{\cos\Theta \big[(k'^2_{3n}/\ell^2+m E_n+m^2)I-i(k'_{3n}/\ell)(k_2\sigma_1-k_1 \sigma_2+(m+E_n)\tan\Theta\sigma_3)\big]\over  (m+E_n)(ik'_{3n}/\ell-m\cos\Theta)}.
\label{rotU2}
\end{align}
Again, taking the correspondence between the obtained rotation operator (\ref{rotU2}) and general formulation Eq.~(\ref{rotoperatorspinspace}), we obtain
\begin{align}
&e^{i\varsigma^{(2)} }\cos({\varphi^{(2)} \over 2})I
={B e^{2ik'_{3n}}\over C}{\cos\Theta (k'^2_{3n}/\ell^2+mE_n+m^2)I\over  (m+E_n)(i(k'_{3n}/\ell)-m\cos\Theta)}
,\\
&e^{i\varsigma ^{(2)}}\sin({\varphi^{(2)} \over 2})\hat {\bm n}^{(2)} \cdot{\bm \sigma}
={B e^{2ik'_{3n}}\over C}{\cos\Theta (k'_{3n}/\ell) \big[ k_2\sigma_1-k_1 \sigma_2+(m+E_n)\tan\Theta\sigma_3\big]\over  (m+E_n)(ik'_{3n}/\ell-m\cos\Theta)}.
\end{align}
Then, it is then straightforward to show that the rotation angle and the rotation axis are given by
\begin{eqnarray}
&&\tan({\varphi^{(2)}\over 2})=-{(k'_{3n}/\ell)\sqrt{(2m^2+2mE_n+k'^2_{3n}/\ell^2
)\tan^2\Theta+{\bm k}^2_\perp\sec^2\Theta}\over  (k'^2_{3n}/\ell^2+mE_n+m^2)},
\label{rotangle2}\\
&&\hat {\bm n}^{(2)}={-k_2\hat x_1+k_1 \hat x_2-(m+E_n)\tan\Theta\hat x_3 \over \sqrt{{\bm k}^2_\perp\sec^2\Theta+(2m^2+2mE_n+k'^2_{3n}/\ell^2
)\tan^2\Theta}},
\label{rotaxis2}
\end{eqnarray}
respectively\footnote{Similar to the result at the first plate, the rotation angle and the rotation axis at the second plate can also be written in the form of $\tan({\varphi^{(2)}/2})=-\tan \Theta k_{3n}\sqrt{2m^2+2mE_n+k'^2_{3n}/\ell^2
+{\bm k}^2_\perp \cot^2\Theta}/ (k'^2_{3n}/\ell^2+m E_n+m^2)$ and $\hat {\bm n}^{(2)}={-k_2\hat x_1+k_1 \hat x_2-(m+E_n)\tan\Theta\hat x_3 /(\tan\Theta \sqrt{2m^2+2mE_n+k'^2_{3n}/\ell^2
+{\bm k}^2_\perp\cot^2\Theta}})$, respectively.  In the case of suppressed perpendicular momentum, they reduce to those of Ref.~\cite{Rohim2021}.}.

From the above results, one can show that $\tan({\varphi^{(1)}\over 2})=\tan({\varphi^{(2)}\over 2})$ and $\hat{\bm n}^{(1)}=-\hat{\bm n}^{(2)}$, which means the reflection at the first plate generates the same rotation angle as at the second plate, but their rotation axis are in the opposite direction. Recalling that at the first and second plates, we have $\chi_R={\cal U}^{(1)}\chi_L$ and $\chi_L={\cal U}^{(2)}\chi_R$, respectively, 
which implies ${\cal U}^{(1)}{\cal U}^{(2)}={\cal U}^{(2)}{\cal U}^{(1)}=I$ 
for the allowed momenta (\ref{DiscreteMomenta}) under 
the reflection in a consistent way.  

We next discuss the energy gap between two states focusing on the case of $m'(\equiv m\ell)\ll 1$ with ${\bm k}_\perp=0$. Using Eq.~\eqref{Discreteenergy}, the energy gap is computed as
\begin{eqnarray}
\Delta E=E_{n+1}-E_n = {\pi\over \ell} - {4 m\cos\Theta\over (4n^2-1)\pi},
\end{eqnarray}
where the value of $\ell$ is fixed. 
The role of the chiral angle appears in the second term and can be understood as the energy gap correction. We note that this correction vanishes for $\Theta=\pi/2$ and $3\pi/2$. 
We can also compute the bound of the energy gap as
\begin{eqnarray}
{\pi\over \ell} - {4 m\over (4n^2-1)\pi} \leq \Delta E \leq {\pi\over \ell}.
\end{eqnarray}
From the above equation, the maximum correction of the lower bound for the energy gap is given by $- {4 m\over (4n^2-1)\pi}$.

\section{Casimir energy}
\label{CasimirEnergy}
In this section, we calculate the Casimir energy of a massive Dirac fermion confined between two parallel plates under the chiral MIT boundary conditions. 
In the presence of this boundary condition, the vacuum energy is given by,
\begin{eqnarray}
E_0=-{L^2\over 2\pi^2}\int^{\infty}_{-\infty}dk_1\int^{\infty}_{-\infty}dk_2\sum^\infty_{n=1}\sqrt{k^2_1+k^2_2+\bigg({k'_{3n}\over \ell}\bigg)^2+m^2},
\label{VacuumE}
\end{eqnarray}
where $L^2$ is the surface area of the plate and the discrete momenta $k'_{3n}\equiv k_{3}\ell$ satisfies the momentum constraint in Eq. (\ref{DiscreteMomenta}). In the above expression, we have used the eigen energies of a Dirac field given in Eq. (\ref{Discreteenergy}). In fact, the vacuum energy in Eq. (\ref{VacuumE}) is divergent. Then, to solve this problem, one can use
the Abel-Plana like summation \cite{Romeo2000} following the procedure used in Refs.~\cite{Bellucci2009, Cruz2018} as follows, 
\begin{eqnarray}
\sum^{\infty}_{n=1} {\pi f(k'_{3n})\over \big(1 -{\sin(2 k'_{3n})\over 2 k'_{3n}}\big)} =-{\pi m'\cos\Theta f(0) \over 2(m'\cos\Theta+1)}+\int^\infty_0 dz f(z) -i\int^\infty_0 dt {f(it)-f(-it)\over {t+m'\cos\Theta\over t-m'\cos\Theta}e^{2t}+1}. 
\label{AbelPlanaLike}
\end{eqnarray}
Considering the denominator of the left-hand side of Eq.~(\ref{AbelPlanaLike}), from the momentum constraint given in Eq.~(\ref{DiscreteMomenta}), we may replace it with the following relation
\begin{eqnarray}
1 -{\sin(2 k'_{3n})\over 2 k'_{3n}}= 1+{m' \cos\Theta\over (m'\cos\Theta)^2+k'^2_{3n}}.
\end{eqnarray}
Then, the above vacuum energy can be rewritten as follows
\begin{align}
E_0=-{L^2\over 2\pi^3 \ell }\int^{\infty}_{-\infty}dk_1\int^{\infty}_{-\infty}dk_2\bigg(-{\pi m'\cos\Theta f(0) \over 2(m'\cos\Theta+1)}+\int^\infty_0 dz f(z) -i\int^\infty_0 dt {f(it)-f(-it)\over {t+m'\cos\Theta\over t-m'\cos\Theta}e^{2t}+1}\bigg),
\label{VacE}
\end{align}
where the function $f(z)$ is defined as
\begin{eqnarray}
f(z)=\sqrt{(k^2_1+k^2_2)\ell^2+z^2+m'^2} \bigg(1+{m' \cos\Theta\over (m'\cos\Theta)^2+z^2}\bigg).
\label{fz}
\end{eqnarray}
In the next step, we can separate the above vacuum energy in Eq.~(\ref{VacE}) into three terms as follows, 
\begin{eqnarray}
E_0=\ell E^{(0)}_0+2 E^{(1)}_0+\Delta E_0. 
\label{threetermvac}
\end{eqnarray}
 $E^{(0)}_0$ in the first term of Eq.~(\ref{threetermvac}) reads
\begin{eqnarray}
E^{(0)}_0=-{L^2\over 4\pi^3}\int^{\infty}_{-\infty}dk_1\int^{\infty}_{-\infty}dk_2 \int^{\infty}_{-\infty}
dk_3 \sqrt{k^2_1+k^2_2+k^2_3+m^2},
\end{eqnarray}
which represents the vacuum energy in the absence of the boundary conditions. The contribution of $E^{(1)}_0$ in the second term of Eq.~(\ref{threetermvac}) is given by
\begin{align}
E^{(1)}_0=-{L^2\over 4\pi^3}\int^{\infty}_{-\infty}dk_1\int^{\infty}_{-\infty}dk_2 \bigg[-{\pi\over 2}\sqrt{k^2_1+k^2_2+m^2}+m\cos\Theta\int^{\infty}_0 dz {\sqrt{z^2+k^2_1+k^2_2+m^2}\over m^2\cos^2\Theta+z^2}\bigg], 
\end{align}
which is the vacuum energy in the presence of one plate only  and does not contribute to the Casimir force. Then, 
the Casimir energy is related to the last term of Eq.~(\ref{threetermvac}), $E_{\rm Cas.}\equiv \Delta E_0$, and is given as 
\begin{eqnarray}
E_{\rm Cas.}={iL^2\over 2\pi^3 \ell }\int^{\infty}_{-\infty}dk_1\int^{\infty}_{-\infty}dk_2\int^\infty_0 dt {f(it)-f(-it)\over {t+m'\cos\Theta\over t-m'\cos\Theta}e^{2t}+1}.
\end{eqnarray}
By using Eq.~(\ref{fz}), the above Casimir energy can be rewritten as,
\begin{eqnarray}
E_{\rm Cas.}&=&{iL^2 \ell \over 2\pi^3 }\int^{\infty}_{-\infty}dk_1\int^{\infty}_{-\infty}dk_2\int^\infty_0 du {u-m\cos\Theta \over (u+m\cos\Theta)e^{2\ell u}+u-m\cos\Theta} 
\nonumber\\
&&
\times  \bigg(1+{m\cos\Theta\over \ell m^2\cos\Theta-\ell u^2}\bigg) \bigg[\sqrt{(iu)^2+k^2_1+k^2_2+m^2}-\sqrt{(-iu)^2+k^2_1+k^2_2+m^2}\bigg],\nonumber\\
\label{Casseparate}
\end{eqnarray}
where we have changed the variable, 
$t=\ell u$. It is convenient to separate the integration over  $u$ in Eq.~(\ref{Casseparate}) into two intervals.  The first interval, i.e. from $0$ to $\sqrt{k^2_1+k^2_2+m^2}$ vanishes, whereas the second interval, i.e. from $\sqrt{k^2_1+k^2_2+m^2}$ to $\infty$ remains.  
Then, the Casimir energy reads
\begin{eqnarray}
E_{\rm Cas.}&=&{-L^2 \over \pi^3 }\int^{\infty}_{-\infty}dk_1\int^{\infty}_{-\infty}dk_2\int^\infty_{\sqrt{k^2_1+k^2_2+m^2}} du \sqrt{u^2-k^2_1-k^2_2-m^2} \nonumber\\
&&~~~~~~~~~~~~~~~\times 
\bigg({ \ell(u-m\cos\Theta)-m\cos\Theta/(m\cos\Theta+u)
\over (u+m\cos\Theta)e^{2\ell u}+(u-m\cos\Theta)}\bigg).
\end{eqnarray}
To further proceed with the above integration, we then use the following formula \cite{Bellucci2009}
\begin{eqnarray}
\int_{-\infty}^{\infty} d{\bm k}_\perp\int^\infty_{\sqrt{{\bm k}^2_\perp+c^2}} dz(z^2-{\bm k}^2_\perp-c^2)^{(s+1)/2} f(z)&=&{\pi^{p/2}\Gamma((s+3)/2)\over \Gamma((p+s+3)/2)}\nonumber\\
&&\times \int_c^\infty dx (x^2-c^2)^{(p+s+1)/2}f(x),
\label{formulaint}
\end{eqnarray}
where $p$  denotes the number of the spatial perpendicular momenta's component (in our case, we have $p=2$).
Then, the Casimir energy becomes
\begin{equation}
E_{\rm Cas.}={-L^2 \over \pi^3}{\pi\Gamma(3/2)\over \Gamma(5/2)}
\int^\infty_m dx  
(x^2-m^2)^{3/2}
\bigg({\ell(x-m\cos\Theta)-m\cos\Theta/(m\cos\Theta+x)
\over (x+m\cos\Theta)e^{2\ell x}+(x-m\cos\Theta)}\bigg).
\end{equation}
We next rewrite the following factor
\begin{eqnarray}
{\ell(x-m\cos\Theta)-m\cos\Theta/(m\cos\Theta+x)
\over (x+m\cos\Theta)e^{2\ell x}+(x-m\cos\Theta)}=-{1\over 2}{d\over dx}\ln{\bigg(1+{x-m\cos\Theta\over x+m\cos\Theta}e^{-2\ell x}\bigg)}
\end{eqnarray}
and introduce a new variable $y=\ell x-m'$. The Casimir energy now reads
\begin{eqnarray}
E_{\rm Cas.}={-L^2 \over \pi^2 \ell^3 } {2\over 3} \int^\infty_0 dy   (y^2+2y m')^{3/2}
{-1\over 2}{d\over dy}\ln{\bigg(1+{y+ m'(1-\cos\Theta)\over y+ m'(1+\cos\Theta)}e^{-2(y+m')}\bigg)}.
\label{intbypart}
\end{eqnarray}
Performing integration by parts in Eq.~(\ref{intbypart}), we have a simpler form of the Casimir energy as
\begin{eqnarray}
E_{\rm Cas.}={-L^2 \over \pi^2 \ell^3 }\int^\infty_0 dy (y+m') \sqrt{y(y+2 m')}\ln{\bigg(1+{y+ m'(1-\cos\Theta)\over y+ m'(1+\cos\Theta)}e^{-2(y+m')}\bigg)}.
\label{intCasimirEnergy}
\end{eqnarray}

\begin{figure}[tbp]
\centering 
\includegraphics[width=.49\textwidth]{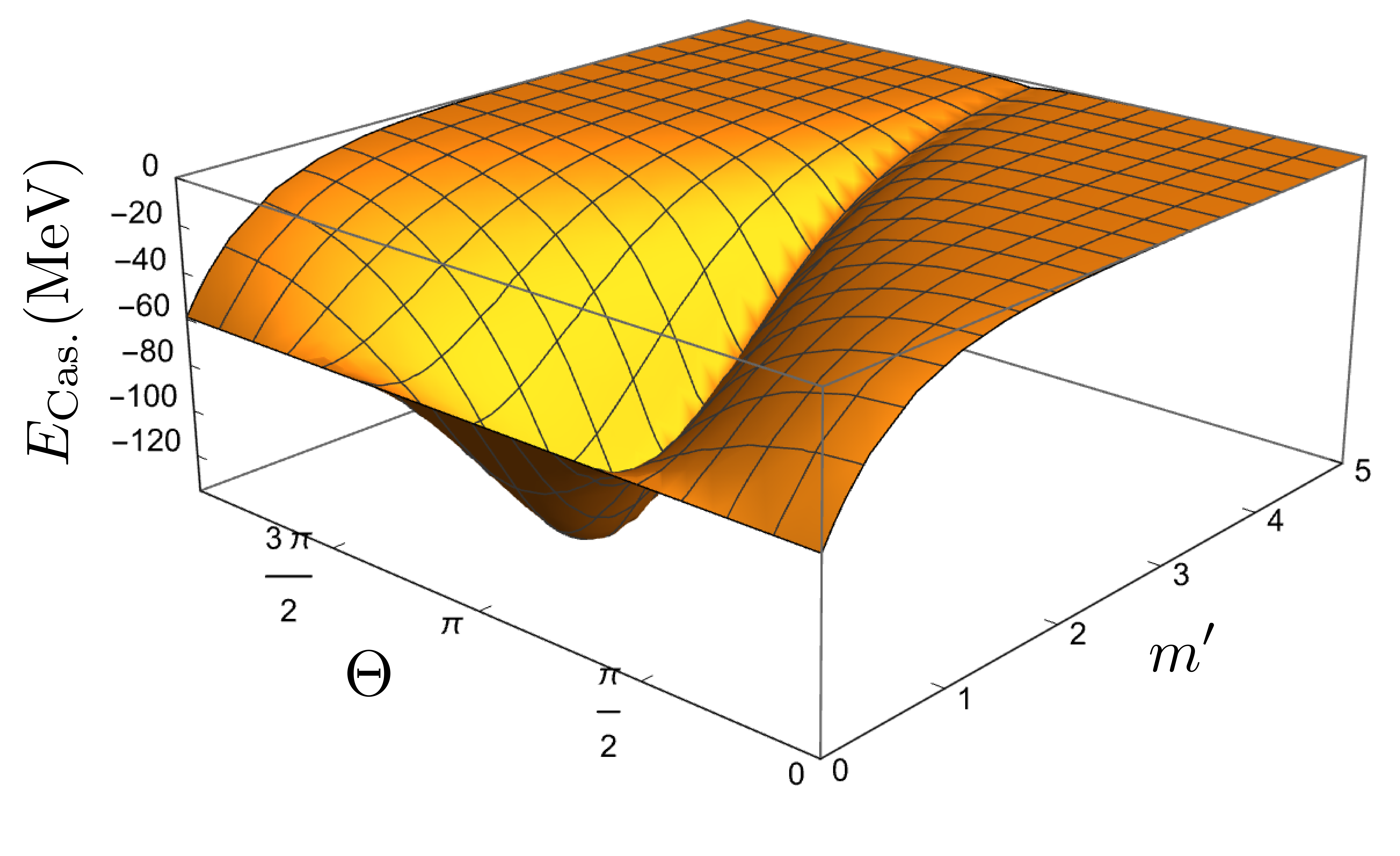}
\hfill
\includegraphics[width=.49\textwidth]{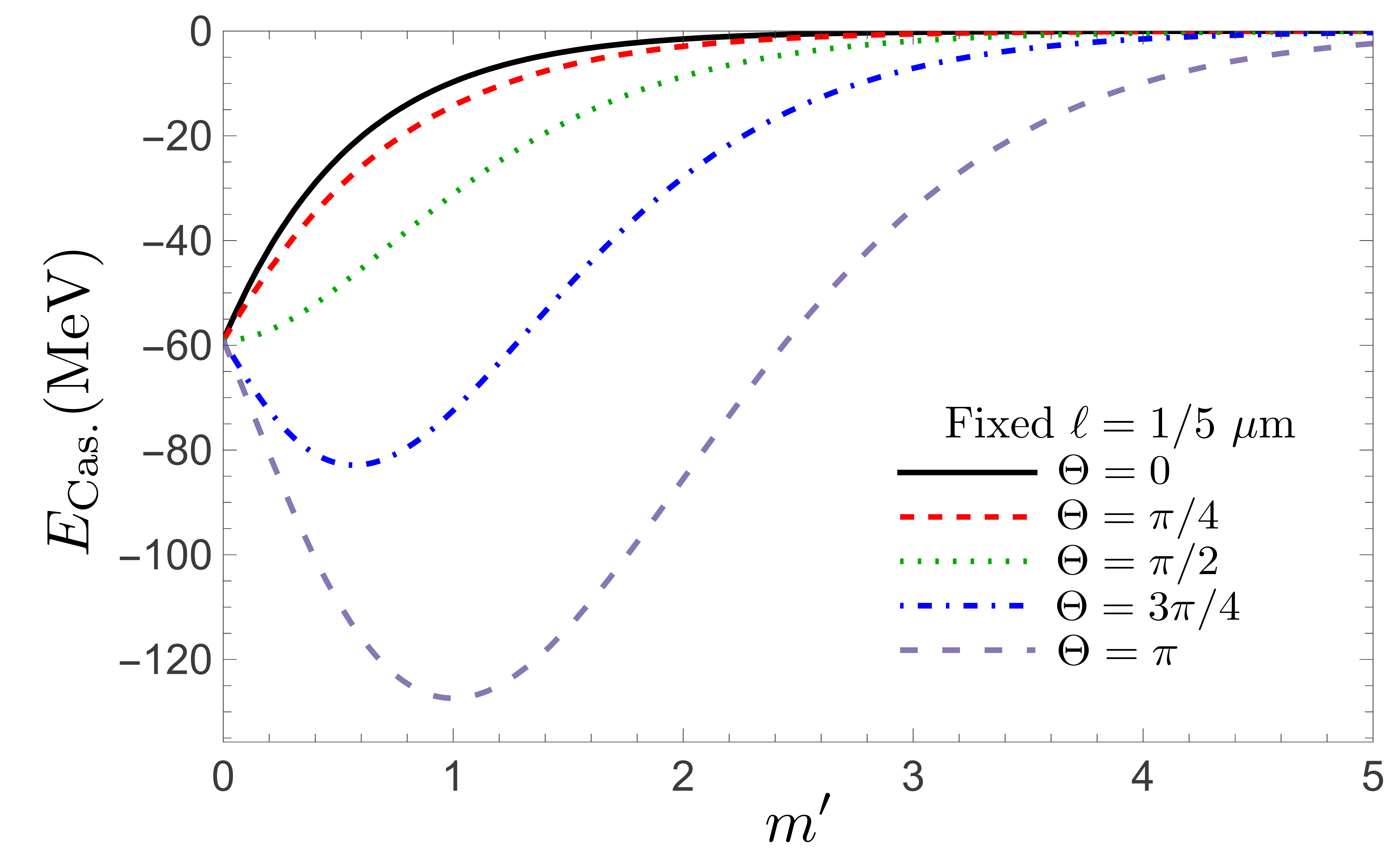}
\caption{\label{CasimirEnergyplot1} The Casimir energy per plate area. The left panel shows the Casimir energy as a function of the chiral angle $\Theta$ and dimensionless parameter $m'(\equiv m\ell)$. In the right panel, the Casimir energy is demonstrated as a function of the parameter $m'$ for various values of the chiral angle. We used a fixed plates' distance of $\ell=1/5~\mu\text{m}$ and plate's size of $L^2=1~\text{cm}^2$. The curves show that in the massless case, the Casimir energy for all chiral angles gives the same value, which is analytically given in Eq.~(\ref{masslessCasimir}). They also show that the chiral angle $\Theta=\pi$ gives maximum contribution. The role of the chiral angle changes depending on the mass $m$ and plates' distance $\ell$.}
\end{figure}
\begin{figure}[tbp]
\centering 
\includegraphics[width=.49\textwidth]{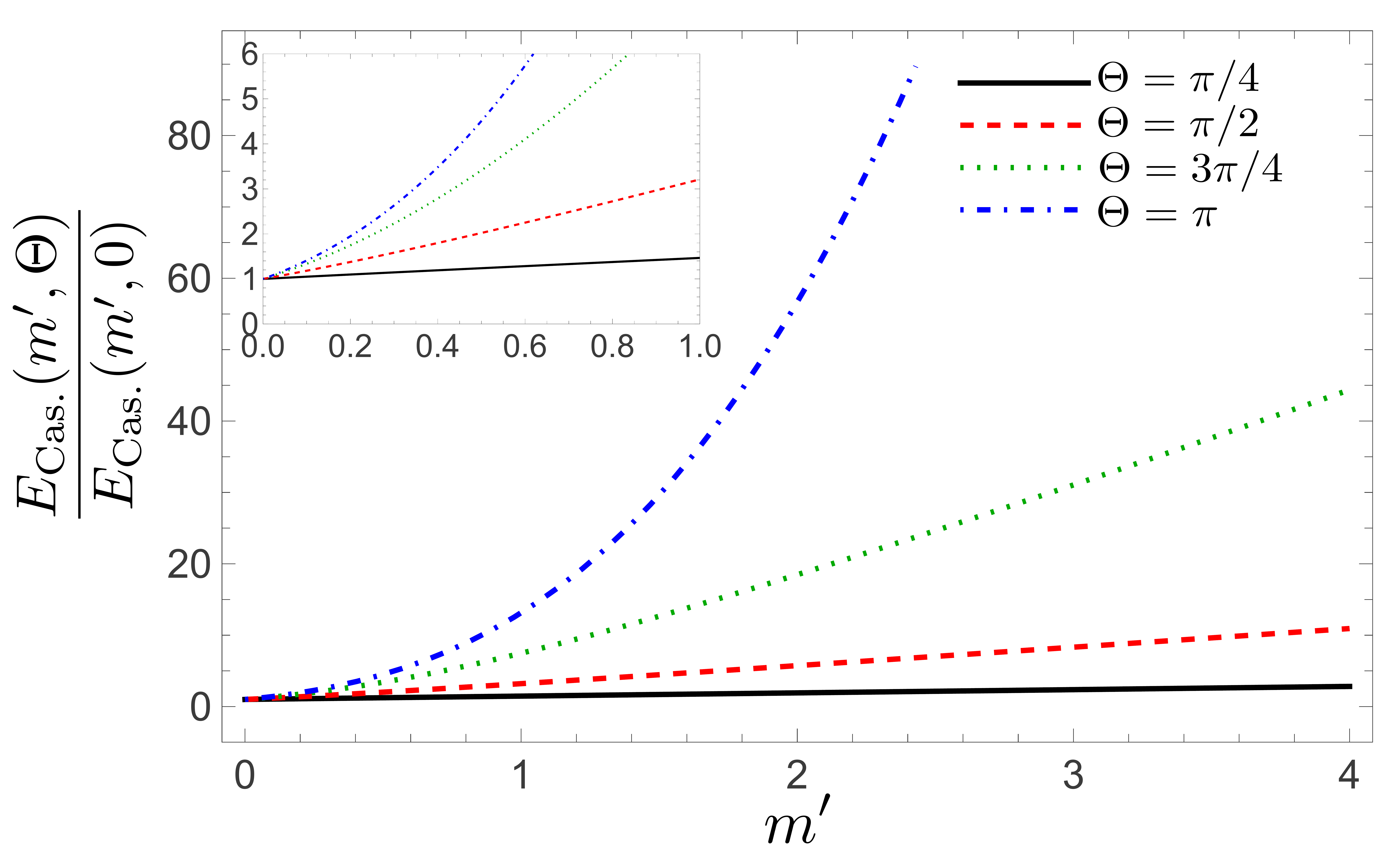}
\hfill
\includegraphics[width=.49\textwidth]{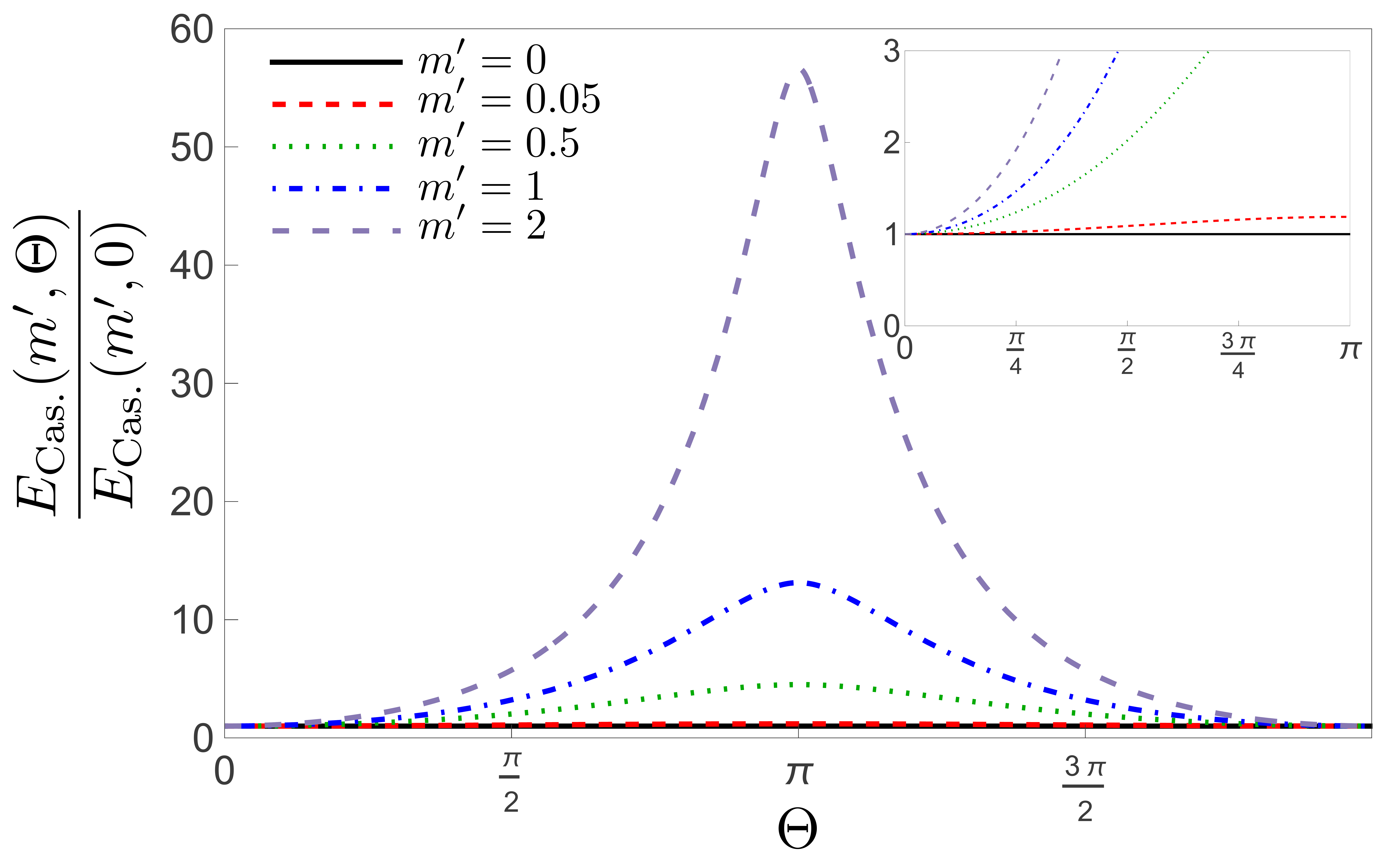}
\caption{\label{CasimirEnergyplot2} The ratio of the Casimir energy for an arbitrary chiral case to the nonchiral case. The left panel  shows  the curves as a function of the parameter $m'$ for several values of the chiral angle. The right panel shows the curves as a function of the chiral angle for the given parameter $m'$.}
\end{figure}
In what follows, we will discuss the Casimir energy based on Eq.~(\ref{intCasimirEnergy}). The left panel of Fig.~\ref{CasimirEnergyplot1} depicts the behavior of the Casimir energy as a function of the chiral angle $\Theta$ and the parameter $m'(\equiv m\ell)$, while the right panel demonstrates the Casimir energy as a function of the parameter $m'$ for several values of the chiral angle. In both panels, we have used the fixed plates' distance.   In general, one can see that the Casimir energy has symmetric shapes with respect to the chiral angle $\Theta=\pi$. 
In the case of the light mass, $m'\ll 1$, the right panel of Fig.~\ref{CasimirEnergyplot1} is approximately given by the linear function, as shown below. While in the case of the heavy mass, $m'\gg 1$, the Casimir energy tends to zero. In the massless case, Fig.~\ref{CasimirEnergyplot1} shows that the Casimir energy gives the same value  for any chiral angle; it is explicitly given by
\begin{eqnarray}
E_{\rm Cas.}={-L^2 \over \pi^2 \ell^3 } \int^\infty_0 dy  y^2\ln (1+e^{-2y})=-{7 L^2\pi^2\over 2880 \ell^3},
\label{masslessCasimir}
\end{eqnarray}
which is consistent with that of Ref.~\cite{KJohnson1975}.  Both right and left panels show that the maximum contribution is given by the chiral angle $\Theta=\pi$, while the minimum contribution is given by $\Theta=0$ (nonchiral case).

Figure~\ref{CasimirEnergyplot2} plots   the ratio of the Casimir energy for an arbitrary chiral case to the nonchiral case. The curves show that the ratio is maximum when the chiral angle takes the value $\Theta=\pi$. This ratio increases as  the parameter $m'$ becomes larger.  
Since the Casimir energy is symmetric as a function of the chiral angle, the ratio $E_{\rm Cas.}(m',\Theta)/E_{\rm Cas.}(m',0)$ is also symmetric with respect to $\Theta=\pi$ (see the right panel of Fig.~\ref{CasimirEnergyplot2}).

\begin{figure}[tbp]
\centering 
\includegraphics[width=.49\textwidth]{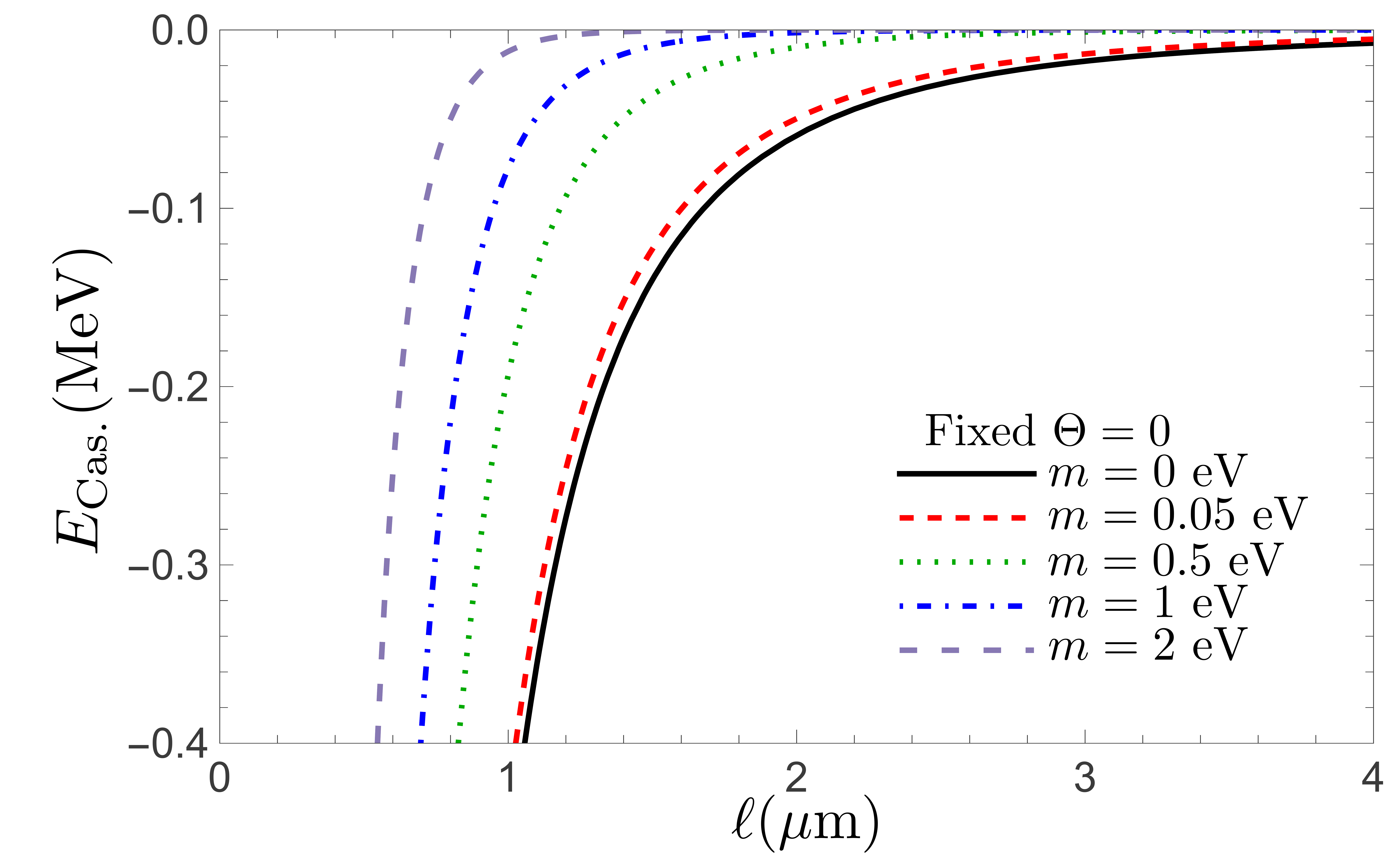}
\hfill
\includegraphics[width=.49\textwidth]{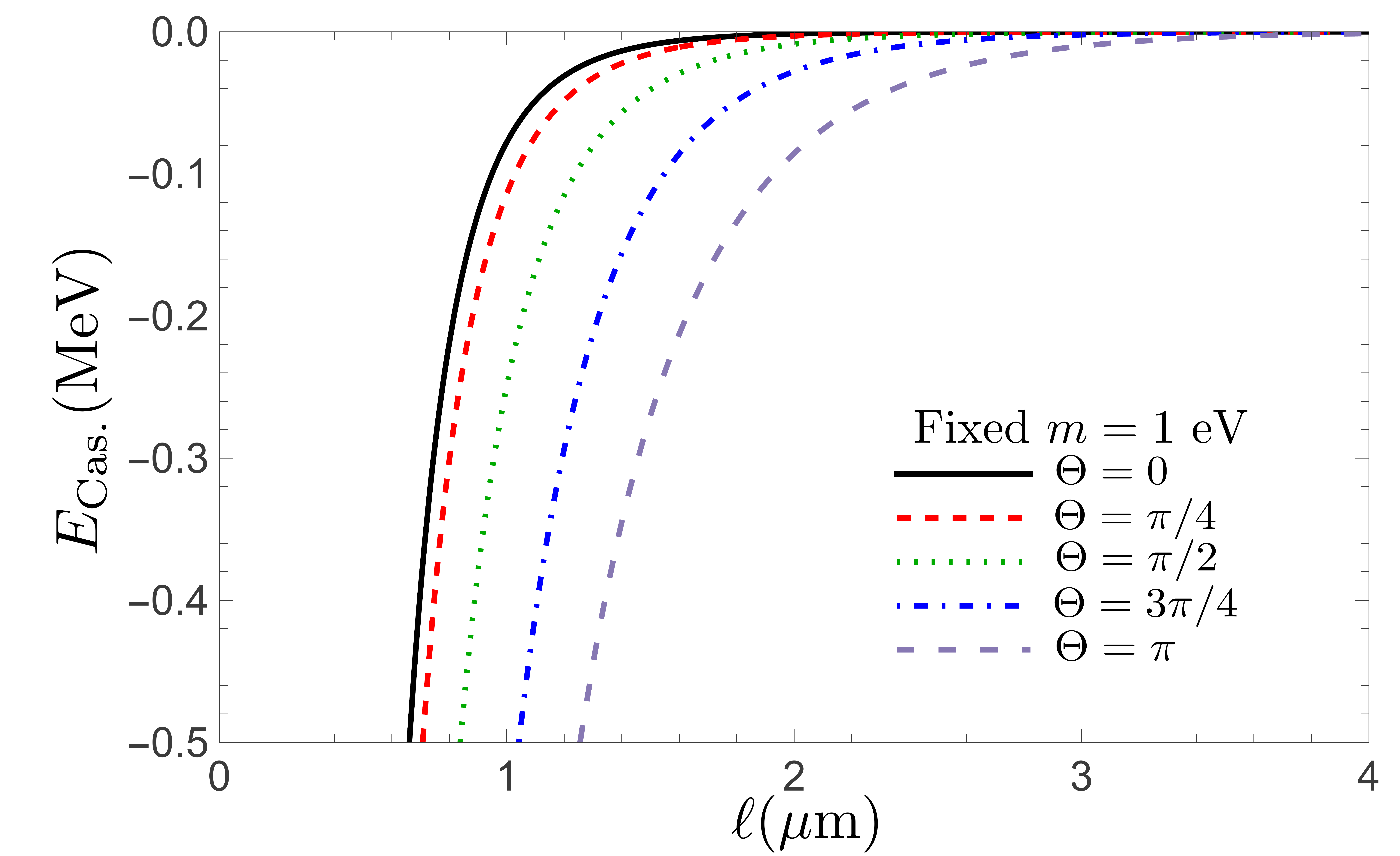}
\caption{\label{CasimirEnergyplot3} The Casimir energy per plate area as a function of $\ell$. The left panel corresponds to the dependence of the Casimir energy on the mass $m$ for the nonchiral case. The right panel corresponds to the dependence of the Casimir energy on the chiral angle $\Theta$ with fixed mass $m=1$ eV and plate's size $L^2=1~\text{cm}^2$.}
\end{figure}

Figure~\ref{CasimirEnergyplot3} plots the Casimir energy as a function of the plates' distance $\ell$. The curves in the left panel show that the Casimir energy of a massive Dirac fermion approaches the massless case as the reduction of its mass. From the right panel, one can see that the Casimir energy in the nonchiral case decays faster than in the chiral case. Meanwhile, the slowest decaying occurs in the case of the chiral angle $\Theta=\pi$.

We next turn to consider the analytical evaluation of the Casimir energy in Eq.~(\ref{intCasimirEnergy}).  
In the case of $m'\ll 1$, 
the Casimir energy  for an arbitrary chiral angle approximately reduces to
\begin{eqnarray}
E_{\rm Cas.}&\simeq&{-L^2 \over \pi^2 \ell^3 } \int^\infty_0 dy \bigg[ y^2\ln (1+e^{-2y})-\bigg({2y {e^{-2y}} (y+\cos\Theta )\over (1+e^{-2y})}-2y\ln (1+e^{-2y})\bigg)m'\bigg]\nonumber\\
&=&-{7 L^2\pi^2\over 2880 \ell^3} \bigg(1-{120 m' \cos\Theta \over 7\pi^2}\bigg).
\label{CasimirEnergymless1}
\end{eqnarray}
Compared to the massless case (\ref{masslessCasimir}), the above expression has an additional term that contributes to the Casimir energy. We also note that this term shows the Casimir energy dependence on the chiral angle. In the case of $m' \gg 1$, the Casimir energy (\ref{intCasimirEnergy}) is approximately given by
\begin{eqnarray}
E_{\rm Cas.} &\simeq& {-L^2 \over \pi^2 \ell^3 } \int^\infty_0 dy \nonumber\\
&&\times \bigg[ \bigg( {\sqrt{2m'^3}y^{1/2} (1-\cos\Theta)\over 1+\cos\Theta}+\sqrt{m'\over 2}{y^{3/2}(5+8\cos\Theta-5\cos^2\Theta)\over 2(1+\cos\Theta)^2}\bigg)e^{-2(y+m')}\bigg]\nonumber\\
&=&{-L^2 m'^{1/2}e^{-2m'} \over 64 (1+\cos\Theta)^2\pi^{3/2} \ell^3 }\bigg[{16 m'\sin^2\Theta} +{3(5+8\cos\Theta-5\cos^2\Theta)}\bigg].
\label{CasimirEnergymgreater1}
\end{eqnarray}
In this case, the Casimir energy 
decays and convergent go to zero as the increase of the parameter $m'$.

 From the above Casimir energy, one can investigate the Casimir force $F_{\rm Cas.}=-\partial E_{\rm Cas.}/\partial \ell$ as well as the Casimir pressure  $P_{\rm Cas.}=F_{\rm Cas.}/L^2$.   From Eqs.~\eqref{CasimirEnergymless1} and \eqref{CasimirEnergymgreater1}, one can obtain the following Casimir pressure, 
\begin{align}
 & P_{\rm Cas.} \nonumber\\
  =&\left\{
\begin{array}{ll}
 -{7 \pi^2 \over 960 \ell^4}\bigg(1-{80 m' \cos\Theta \over 7\pi^2}\bigg), ~~ \text{for $m'\ll 1$},\\
  -{m'^{1/2}e^{-2m'} \over 64 (1+\cos\Theta)^2\pi^{3/2} \ell^4 }\bigg[{ 8( 4m'+3)m'\sin^2\Theta} +(6 m'+{15\over 2})(5+8\cos\Theta-5\cos^2\Theta)\bigg],~~ \text{for $m'\gg 1$}.
\end{array}
\right.
\label{CasimirPressure}
\end{align}
The above Casimir pressures with respect to the chiral angle have a similar behavior to those of the Casimir energy. In other word, the attractive Casimir force in the chiral case is always stronger than that in the nonchiral case.

\section{Summary 
\label{Summary}}

We have studied the behavior of a massive Dirac fermion confined between two parallel chiral plates. In our setup, the plates are represented by chiral MIT boundary conditions \cite{Lutken1984}, where the condition of vanishing probability current density at the boundary surfaces is satisfied for arbitrary chiral angles. The Dirac field inside the confinement area consists of two-component fields associated with their spin orientations.  We discussed the general discrete momenta and the changes in the spin orientations under boundary conditions.  The result shows that only momentum $k_3$ (parallel component to the normal plate surface) is discretized depending on the mass, plates' distance, and the chiral angle. We also found that, in the case of non-zero perpendicular momenta, the spin orientation changes for an arbitrary chiral angle. In the case of suppressed perpendicular momenta, we found that these features reduce to those of the previous work \cite{Rohim2021} for a confinement system in a one-dimensional box. 

We also discussed the energy gap between two states in the case of the light mass ($m'\ll1$). The result shows that the effect of the chiral angle appears as the correction of the energy gap in the massless case. In this context, we may address such an effect on the electron transport in materials such as graphene nanoribbons \cite{MYHan2007, YMLin2008, Han2010}. Since the chiral angle is coupled to the mass, the correction term with the chiral angle is quite small depending on the mass value. Our present study could also be applicable to nanotubes, where one can consider a confinement system of a Dirac field between two parallel plates with compactified dimensions \cite{Bellucci2009}. The detailed analysis of such an application is beyond the scope of the present study and will be presented elsewhere.

We have also investigated the effect of the chiral angle on the Casimir energy of a massive fermion field. To obtain the Casimir energy, we calculated the vacuum energy in the presence of the boundary condition. Unfortunately, this vacuum energy is divergent. To solve this issue, we adopt the Abel-Plana like summation \cite{Romeo2000}, as previously used in Refs.~\cite{Cruz2018, Bellucci2009}. 
The obtained vacuum energy consists of three parts (see Eq.~\eqref{threetermvac}), i.e., (i) the vacuum energy in the absence of the boundary conditions, (ii)  the vacuum energy in the presence of a single plate, (iii) the Casimir energy. We notice that the second term of the vacuum energy in Eq.~(\ref{threetermvac}) for a single plate is not relevant to the Casimir force. The Casimir energy is defined by taking the differences between the vacuum energy of a Dirac field in the presence of the boundary conditions to that in the absence of one. We investigated the behavior of the Casimir energy as well as its attractive force using numerical analysis. The result shows that the Casimir energy of a massive fermion field can be written as a function of the chiral angle. It is symmetric with the maximum contribution occuring at $\Theta=\pi$; the Casimir energy in the chiral case is always higher than that in the nonchiral case. 

In addition to the above results, we also found that the behavior of the Casimir energy depends on the mass of the Dirac fermion. In the analysis, we investigated two approximation cases, i.e. light and heavy masses.  In the case of heavy fermion mass, the Casimir energy converges to zero as the mass increases, whereas in the case of light fermion mass, the Casimir energy converges to that of the massless fermion as the mass decreases. We found that, in both cases, the roles of the chiral angle become weaker.  For the case of the massless fermion, the Casimir energy gives the same value for all chiral angles. In this case, we recover the expected result by Ref.~\cite{KJohnson1975}. For future work, it will be interesting to study a similar setup by including a background such as a magnetic field (c.f., Ref.~\cite{Sitenko:2014kza}).

\section*{Acknowledgements}
{This work was started when A.~R. stayed at Kyushu University, Japan as an Academic Researcher (postdoctoral fellow) and was completed during he was the postdoctoral program at the National Research and Innovation Agency (BRIN), Indonesia. 
 A.~R. would like to thank the Theoretical Astrophysics Laboratory of Kyushu University for their kind hospitality.
We also thank A.~N.~Atmaja for fruitful discussions.
 }

\appendix
 
\section{Complementary derivations for discrete momenta} 
\label{detaildiscretemomenta}

 Two-component Dirac spinors $\chi_{R(L)}$ can be decomposed  as
 \begin{eqnarray}
 \chi_{R(L)}=  \begin{pmatrix}
\alpha_{R(L)}\\
\beta_{R(L)}
\end{pmatrix}
.
\label{componenttwospinor}
\end{eqnarray}
 The relation in Eq.~(\ref{relation01}) can then be rewritten in a more explicit way as follows
\begin{eqnarray}
 &&B= -C {(-ik_3(1+\sin\Theta)-(m+E)\cos\Theta)\alpha_L+i(k_1-ik_2)(1+\sin\Theta)\beta_L
 \over (ik_3(1+\sin\Theta)-(m+E)\cos\Theta)\alpha_R+i(k_1-ik_2)(1+\sin\Theta)\beta_R},
 \label{relation0}
 \label{rel11}\\
 &&B= -C {i(k_1+ik_2)(-1+\sin\Theta)\alpha_L+(ik_3(-1+\sin\Theta)-(m+E)\cos\Theta)\beta_L
 \over 
 i(k_1+ik_2)(-1+\sin\Theta)\alpha_R +(-ik_3(-1+\sin\Theta)-(m+E)\cos\Theta)\beta_R}. 
  \label{rel12}
\end{eqnarray}
In a similar way, from Eq.~(\ref{relationL1}), we have the relation between coefficients $B$ and $C$  at the second plate $x_3=\ell$ as follows 
\begin{align}
 &B= -C e^{-2ik_3\ell}{(-ik_3(-1+\sin\Theta)-(m+E)\cos\Theta)\alpha_L+i(k_1-ik_2)(-1+\sin\Theta)\beta_L
 \over (ik_3(-1+\sin\Theta)-(m+E)\cos\Theta)\alpha_R+i(k_1-ik_2)(-1+\sin\Theta)\beta_R},
  \label{rel21}\\
 &B= -C e^{-2ik_3\ell} {i(k_1+ik_2)(1+\sin\Theta)\alpha_L+(ik_3(1+\sin\Theta)-(m+E)\cos\Theta)\beta_L
 \over 
 i(k_1+ik_2)(1+\sin\Theta)\alpha_R +(-ik_3(1+\sin\Theta)-(m+E)\cos\Theta)\beta_R}.  \label{rel22}
\end{align}
By equating Eqs.~(\ref{rel11}) with (\ref{rel21}), we have
\begin{eqnarray}
(A_1\alpha_R+A_2\beta_R)\alpha_L+(A_3\alpha_R+A_4\beta_R)\beta_L=0,
\label{boundaryrel1}
\end{eqnarray}
where
\begin{eqnarray}
A_1 &=& e^{-2 ik_3 \ell}
  \cos\Theta \big[2 i (1 + e^{2 i k_3 \ell}) k_3 (E + m) \nonumber
  \\
  &&+ (-1 + e^{2 i k_3 \ell}) (E - k_3 + m) (E + k_3 + m) \cos\Theta\big],
  \\
A_2&=&e^{-2 i k_3 \ell}  \cos\Theta (k_1 - i k_2)\big[(1 - e^{2 i k_3 \ell}) k_3 \cos\Theta \nonumber
  \\
  &&- 
   i (E + m) (-1 - 
      \sin\Theta+ e^{2 i k_3 \ell} (-1 + \sin\Theta) )\big],
  \\
A_3&=&e^{-2 i k_3 \ell}  \cos\Theta (k_1 - 
   i k_2) \big[(-1 + e^{2 i k_3 \ell}) k_3 \cos\Theta \nonumber
  \\
  &&+ 
   i (E + m) (-1 + \sin\Theta - e^{2 i k_3 \ell} (1 + \sin\Theta))\big],
  \\
 A_4&=& \cos^2\Theta (k_1 - i k_2)^2 (1-e^{-2ik_3\ell}).
\end{eqnarray}
Taking same procedure as above for Eqs.~(\ref{rel12}) and (\ref{rel22}), we have
\begin{eqnarray}
(B_1\alpha_R+B_2\beta_R)\alpha_L+(B_3\alpha_R+B_4\beta_R)\beta_L=0,
\label{boundaryrel2}
\end{eqnarray}
where
\begin{eqnarray}
&&B_1= {(k_1+ik_2)^2\over (k_1-ik_2)^2}A_4,\\
&&B_2={k_1+i k_2\over k_1-ik_2}A_2,\\
&&B_3={k_1+i k_2\over k_1-i k_2}A_3,\\ 
 &&B_4=A_{1}.
\end{eqnarray}
The relation given in Eqs. (\ref{boundaryrel1}) and (\ref{boundaryrel2}) can be written simultaneously in the form of multiplication between $2\times 2$ and $2\times 1$ matrices as follows  
\begin{eqnarray}
 \begin{pmatrix}
(A_1 \alpha_R+A_2\beta_R) ~&~ (A_3 \alpha_R+A_4\beta_R)\\
(B_1 \alpha_R+B_2\beta_R) ~&~ (B_1 \alpha_R+B_4\beta_R)
\end{pmatrix}
 \begin{pmatrix}
\alpha_L\\
\beta_L
\end{pmatrix}
=0.
\label{matrixBC}
\end{eqnarray}
The nontrivial values of $\alpha_L$ and $\beta_L$ require the $2\times 2$ matrix determinant in Eq.~(\ref{matrixBC}) to vanish, which leads to the following condition
\begin{eqnarray}
D_1\alpha^2_R+D_2\alpha_R\beta_R+D_3\beta^2_R=0,
\label{determinantBC}
\end{eqnarray}
where
\begin{align}
D_1=& A_1B_3-A_3B_1\nonumber\\
=&-2 e^{-4 ik_3 \ell}
  \cos^2\Theta (E+m) \big[i (1 + e^{2 i k_3 \ell}) k_3+ (-1 + e^{2 i k_3 \ell})m \cos\Theta\big]\nonumber\\
  &~\times (k_1+ik_2) \big[-(-1+e^{2 i k_3 \ell})k_3\cos\Theta+i(E+m)(1-\sin\Theta+e^{2 i k_3 \ell}(1+\sin\Theta))\big],\\
D_2=&A_1B_4+A_2B_3-A_3B_2-A_4B_1\nonumber\\
=&4 e^{-4 ik_3 \ell}
  \cos^2\Theta (E+m) \big[i (1 + e^{2 i k_3 \ell}) k_3+ (-1 + e^{2 i k_3 \ell})m \cos\Theta\big]\nonumber\\
  &~\times \big[i (1 + e^{2 i k_3 \ell}) k_3 (E+m)+ (-1 + e^{2 i k_3 \ell})(E^2+mE-k^2_3) \cos\Theta\big],\\
D_3=& A_2B_4-A_4B_2\nonumber\\
=&2 e^{-4 ik_3 \ell}
  \cos^2\Theta (E+m) \big[i (1 + e^{2 i k_3 \ell}) k_3+ (-1 + e^{2 i k_3 \ell})m \cos\Theta\big]\nonumber\\
  &~\times (k_1-ik_2) \big[-(-1+e^{2 i k_3 \ell})k_3\cos\Theta-i(E+m)(-1-\sin\Theta+e^{2 i k_3 \ell}(-1+\sin\Theta))\big].
\end{align}
From the previous expressions, we can see that $D_1,~D_2,~D_3$ have the same factor as follows
\begin{eqnarray}
2 e^{-4 ik_3 \ell}
  \cos^2\Theta (E+m) \big[i (1 + e^{2 i k_3 \ell}) k_3+ (-1 + e^{2 i k_3 \ell})m \cos\Theta\big].
\end{eqnarray}
To satisfy the condition in Eq.~(\ref{determinantBC}) for arbitrary chiral angle $\Theta$, the above factor must vanish 
\begin{eqnarray}
i (1 + e^{2 i k_3 \ell}) k_3+ (-1 + e^{2 i k_3 \ell})m \cos\Theta=0,
\end{eqnarray}
which leads to the allowed condition for discrete momenta in Eq.~(\ref{DiscreteMomenta}).

\let\doi\relax

\end{document}